\documentclass[draftnofoot, 12pt, onecolumn]{IEEEtran}

\usepackage{setspace}
\usepackage{graphicx,dblfloatfix}
\usepackage{times}
\usepackage[cmex10]{amsmath}
\usepackage{amssymb}
\usepackage{array}
\usepackage{mathrsfs}
\usepackage{epsf}
\usepackage[tight,footnotesize]{subfigure}

\newcolumntype{M}{>{\centering\arraybackslash}m{0.12\textwidth}}

\pdfoutput=1

\begin{document}

\title{Fog Computing: Focusing on Mobile Users at the Edge}
\author{
\IEEEauthorblockN{Tom~H.~Luan\IEEEauthorrefmark{1}, Longxiang~Gao\IEEEauthorrefmark{1}, Zhi~Li\IEEEauthorrefmark{3}, Yang~Xiang\IEEEauthorrefmark{1}, Guiyi~We\IEEEauthorrefmark{2}, and Limin~Sun\IEEEauthorrefmark{3}} \\
\IEEEauthorblockA{\IEEEauthorrefmark{1} School of Information Technology, Deakin University, Melbourne Burwood, VIC 3125, Australia \\
        \IEEEauthorrefmark{2} School of Computer Science and Information Engineering, Zhejiang Gongshang University, Zhejiang, China \\
        \IEEEauthorrefmark{3} Institute of Information Engineering, Chinese Academy of Sciences, Beijing, China}
}
\maketitle

\begin{abstract}
With smart devices, particular smartphones, becoming our everyday companions, the ubiquitous mobile Internet and computing applications pervade people's daily lives. With the surge demand on high-quality mobile services at anywhere, how to address the ubiquitous user demand and accommodate the explosive growth of mobile traffics is the key issue of the next generation mobile networks. The Fog computing is a promising solution towards this goal. Fog computing extends cloud computing by providing virtualized resources and engaged location-based services to the edge of the mobile networks so as to better serve mobile traffics. Therefore, Fog computing is a lubricant of the combination of cloud computing and mobile applications. In this article, we outline the main features of Fog computing and describe its concept, architecture and design goals. Lastly, we discuss some of the future research issues from the networking perspective.
\end{abstract}

\section{Introduction}

\label{section: introduction}

Our networking today is shaped by two obvious trends.

\begin{itemize}
\item \textbf{Cloud-based Internet}: Cloud computing has already evolved as the key computing infrastructure for Internet with full-fledged services encompassing not only contents but also communications, applications and commerce. As reported, around $90\%$ of global Internet users are now relying on the services provided by cloud, either directly through consumer services or indirectly through their service provider's reliance upon different commercial clouds.

\item \textbf{Proliferation of mobile computing}: Since 2011, the smartphone shipment worldwide has overtook that of PCs. To date, the smartphone penetration in U.S. has reached 80\%. As predicted by Cisco, the average connected devices per person will reach 6.58 in 2020. With various smart devices bringing strong computing and communication power to the palm of people's hand at anywhere, a variety of mobile computing applications, \emph{e.g.}, virtual reality, sensing and navigation, have emerged and resulted in the fundamental changes in the pattern that people live.
\end{itemize}

With cloud computing becoming the overarching Internet approach for information
storage, retrieval and management, and mobile devices becoming the major
outlets of service applications, the successful integration of cloud
computing and mobile devices therefore represents the key task for the next
generation network. This however faces several fundamental challenges:
\begin{itemize}
\item \textbf{Agility of services}: Unlike traditional PC users which typically ask for common Internet applications, such as emails and Internet surfing, mobile users wearing different kinds of mobile devices may request for highly diverse applications, \emph{e.g.}, ehealthcare, Internet-of-Things (IoT) applications, which adapt with their locations and the environment. The centralized cloud can hardly manage the diverse service requests from billions of nomadic mobile users.
\item \textbf{Real-time response}: With mobile devices limited in resources by nature, mobile applications typically need to outsource their computation jobs to the cloud, and expect real-time response. The emerging wearable devices, \emph{e.g.}, Google glasses and Microsoft Hololens, would rely even more intensively on cloud to support the real-time sensing and data processing.
\item \textbf{Long-thin connection}: The enjoyable high-rate data exchange between cloud and mobile is fundamental to support the resource-hungry mobile applications. This however is still impractical which is impeded by the long-thin connections between mobile users and remote cloud. The high wireless bandwidth cost is also daunting to mobile users. As reported by a recent Cisco survey, even of poor reliability and security, WiFi is still favored by 50$\%$ of smartphone users and over 80$\%$ tablets, laptop and eReader users than cellular networks due to its low cost.
\end{itemize}


To overcome above issues between cloud and mobile applications, Fog computing has recently emerged as a more practical solution to enable the smooth convergence between cloud and mobile for content delivery and real-time data processing \cite{stojmenovic2015overview}. The term\textquotedblleft Fog computing" was first introduced by Cisco in 2012 \cite{bonomi2012fog}. Similar systems typically known as edge computing, such as Cyber Foraging \cite{cyber_forage}, Cloudlets \cite{cloudlets} and mobile
edge computing \cite{portal2014mobile}, can date back to early 2000.

The idea of Fog computing is by placing light-weight cloud-like facility at the proximity of mobile users; the Fog therefore can serve mobile users with a direct short-fat connection as compared to the long-thin mobile cloud connection. More importantly, as deployed at localized sites, Fog computing can provide customized and engaged location-aware services which are more desirable to mobile users.

While been actively pursued recently, Fog computing is still new and lack of a standardized definition. This paper provides a view of Fog computing from the networking perspective with the goal to shape the key features of Fog computing and identify its main design goals and open research issues towards an efficient mobile networking system. In the rest part of this article, we unfold our journey by first describing the basic system architecture of Fog computing and showcase some exemplary application scenarios. After that, we discuss on the fundamental motivation behind the Fog computing and its comparisons with
existing related networking paradigms. Lastly, we discuss on the potential
research directions and close this article with concluding remarks.

\section{System Architecture}

\begin{figure}[t]
\centering
\includegraphics[width=.7\textwidth]{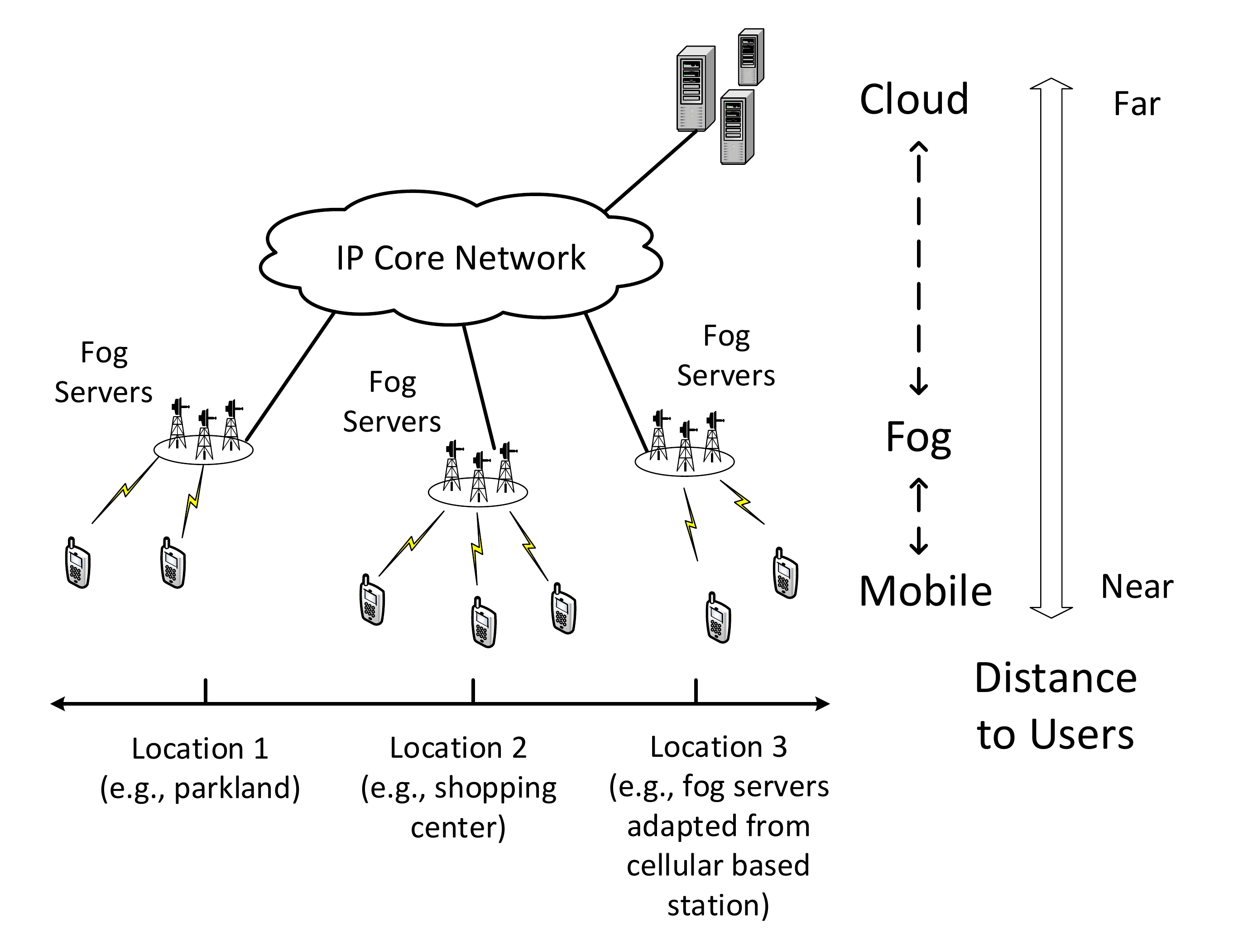}
\caption{Fog computing architecture}
\label{fig: fog architecture}
\end{figure}

Fig.~\ref{fig: fog architecture} illustrates a brief structure of Fog
computing. The Fog computing extends cloud computing by introducing an
intermediate Fog layer between mobile devices and cloud. This accordingly leads
to a three-layer Mobile-Fog-Cloud hierarchy.

The intermediate Fog layer is composed of geo-distributed Fog servers which are deployed at the local premises of mobile users, \emph{e.g.}, parks, bus terminals, shopping
centers, \emph{etc}.. A Fog server is a virtualized device with build-in
data storage, computing and communication facility; the purpose of Fog
computing is therefore to place a handful of compute, storage and
communication resources in the close proximity of mobile users, and accordingly provide fast-rate services to mobile users via the local short-distance high-rate wireless connections. A Fog server can be adapted from existing network components, \emph{e.g.}, a cellular base station, WiFi access point or femtocell router by upgrading the
computing and storage resources and reusing the wireless interface. A Fog server can be static at a fixed location, \emph{e.g.}, inside a shop installed similar
as a WiFi access point, or mobile placed on a moving vehicle as the
Greyhound \textquotedblleft BLUE" system \cite{greyhound}. Some exemplary
use cases will be discussed in the next section.

The role of Fog servers is to bridge the mobile users and cloud. On one
hand, Fog servers directly communicate with mobile users through single-hop
wireless connections using the off-the-shelf wireless interfaces, such as
WiFi, cellular or Bluetooth\footnote{Apple has released iBeacon framework in iOS to support Fog computing services using Bluetooth.}. With cloud-like resources, a Fog server is able to independently provide pre-defined application services to mobile users in its wireless coverage without the assistances of other Fog servers or remote cloud. On the other hand, the Fog servers can be connected to the cloud over Internet
so as to leverage the rich computing and content resources of cloud.

To summarize, the Fog computing is to deploy the virtualized cloud-like device more close to mobile users, and therefore the Fog is interpreted as \textquotedblleft the
cloud close to the ground" \cite{bonomi2012fog}. In what follows,
we will discuss on the rationale of Fog computing and exhibit some
exemplary application scenarios.

\section{Why Fog? Put Service Close to Consumer}

The motivation of Fog computing is to place the contents and application services as close as possible to their consumers. In particular, we argue that the Fog
computing is dedicated to serving mobile users by addressing the shortage of
location-awareness of cloud computing.

Specifically, in contrast to traditional PC users, mobile users have predictable
service demands subject to their locations \cite{bacstuug2014living}. For example, a mobile user in a shopping center tends to be interested in the local sales, open hour,
restaurants and events inside the attended shopping center; such information
become useless once he/she leaves the shopping center. In another example, a
visitor to a new city would seek for the information on the places of
interest, news and weather conditions of that city, while unlikely to be
interested in such information of other places.

\begin{figure*}[tp!]
\subfigure[Retrieving the flyer from the cloud]{
    \label{fig:example:a}     \begin{minipage}[b]{0.5\textwidth}
      \centering
      \includegraphics[width=2.8in]{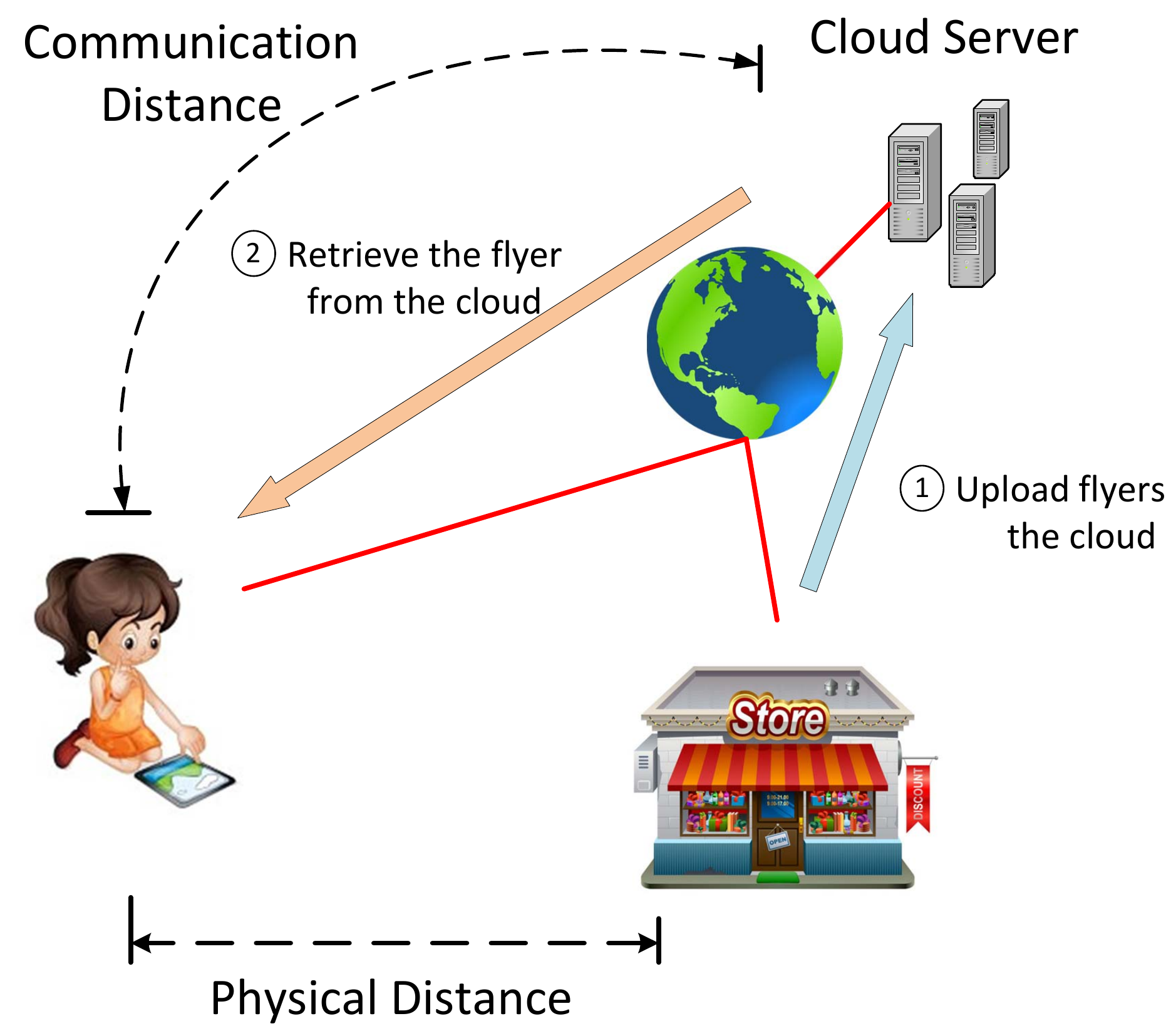}
    \end{minipage}}
\subfigure[Retrieving the flyer from the fog]{
    \label{fig:example:b}     \begin{minipage}[b]{0.5\textwidth}
      \centering
      \includegraphics[width=2.8in]{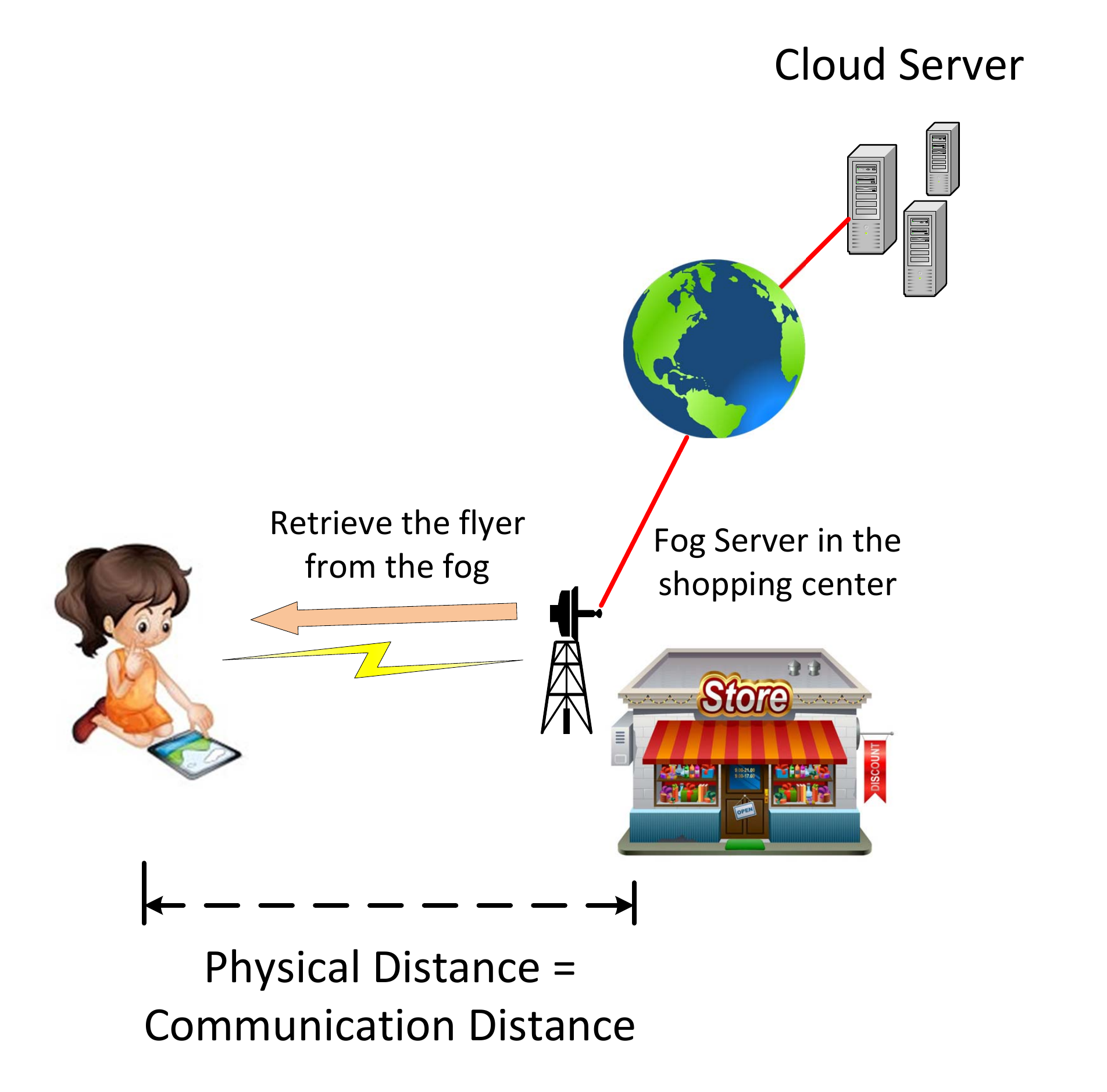}
    \end{minipage}} 
\caption{Example: download the flyer of a nearby store}
\label{fig:example}
\end{figure*}

Cloud computing provides a central portal of information, but is lack of
location-awareness. Such model is suitable for indoor PC users with a
high-rate wired connection; it is however not only costly
for mobile users using expensive cellular bandwidth, but also inconvenient as mobile users have to punch fingers over a slim touch screen to dig in a global pool of information for specific local contents. As a toy example shown in Fig.~\ref{fig:example:a}, assuming that a mobile user inside a shopping center is to download the localized store flyers within the shopping center. To do this using the traditional cloud-based Internet, the stores may need
to first upload their flyers to a remote cloud server, and then direct
mobile users to retrieve the contents through a long-distance link from the cloud,
although the store and mobile user are physically close to each other.

Fog computing overcome this issue by providing engaged localized services
subject to the specific deployment sites. In the same example in Fig.~\ref{fig:example:b}, a Fog server can be deployed inside the shopping center and
pre-cache the localized contents. The mobile users can therefore enjoy
high-rate local connections without the need to search over cloud.

As indicated in this example, the Fog computing brings two immediate
advantages:

\begin{itemize}
\item \textbf{Enhanced service quality to mobile users}: As compared to cloud, Fog computing can provide enhanced service quality with much increased data rate and reduced service latency and response time. Moreover, by downloading through local connections without going through the backbone network, the users can benefit from the reduced bandwidth cost.

\item \textbf{Enhanced efficiency to network}: Fog computing avoids the back-and-forth traffic between cloud and mobile users. This not only saves the backbone bandwidth, but also significantly reduces the energy consumption and carbon footprint of core networks, and therefore represents a promising approach towards the sustainable development of networking.
\end{itemize}

\begin{table*}[tbp]
\caption{Comparison of Fog Computing and Cloud Computing}
\label{tab: compare}\centering
\begin{tabular}{M|m{0.39\textwidth}|m{0.39\textwidth}}
\hline\hline
& \hspace{0.14\textwidth} \textbf{Fog Computing} & \hspace{0.14\textwidth} \textbf{Cloud Computing} \\ \hline\hline
Target User & Mobile users & General Internet users. \\ \hline
Service Type & Limited localized information services related to specific
deployment locations & Global information collected from worldwide \\ \hline
Hardware & Limited storage, compute power and wireless
interface & Ample and scalable storage space and compute power \\ \hline
Distance to Users & In the physical proximity and communicate through
single-hop wireless connection & Faraway from users and communicate through
IP networks \\ \hline
Working Environment & Outdoor (streets, parklands, \emph{etc}.) or indoor
(restaurants, shopping malls, \emph{etc}.) & Warehouse-size building with air conditioning systems\\
\hline
Deployment & Centralized or distributed in reginal areas by local business
(local telecommunication vendor, shopping mall retailer, \emph{etc}.) & Centralized and
maintained by Amazon, Google, \emph{etc}.
 \\ \hline\hline
\end{tabular}
\end{table*}

\section{Exemplary Applications}

Fog computing, composed of geo-distributed Fog servers, targets to deliver the localized and location-based services. In what follows, we showcase some examples of Fog computing implementation from this perspective. Some other use cases of Fog computing in IoT applications are described in \cite{ciscofogwhite, stojmenovic2015overview}.

\paragraph{Shopping Center}

A number of Fog servers can be deployed at different levels of a multi-floor
shopping center, and collectively form an integrated localized Fog computing information
system. The Fog servers at different levels can pre-cache floor-specific contents, such as the layout and ads of stores on the current floor, and providing accurate location-based applications including indoor navigation, ads distribution and feedback collections to mobile users through WiFi.

\paragraph{Senary Park}

The Fog computing system can be deployed in a senary park to provide
localized tourism services. For instance, Fog servers can be deployed at the
entrance or important locations inside the park. The Fog server at the
park entrance can pre-cache information including map and tourist guide; other Fog servers at different locations inside the park can be incorporated with sensor networks for environment monitoring and provide navigation and alert information to tourists.

\begin{figure}[t]
\centering
\includegraphics[width=.7\textwidth]{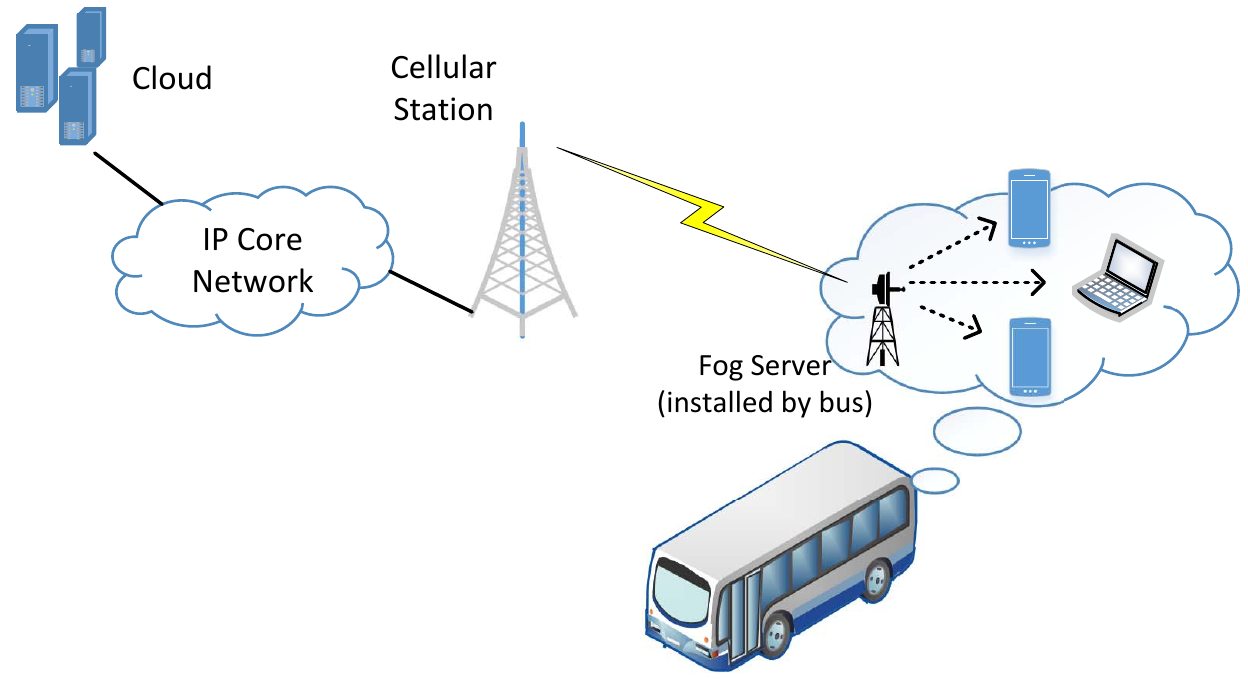}
\caption{On-board Fog computing system}
\label{fig: bus}
\end{figure}

\paragraph{Inter-state Bus}

Greyhound has launched ``BLUE" \cite{greyhound}, an on-board Fog computing
system over inter-state buses for entertainment services. As an example
illustrated in Fig.~\ref{fig: bus}, a Fog server can be deployed inside the
bus and provides on-board video streaming, gaming and social networking
services to passengers using WiFi. The on-board Fog server connects to the
cloud through cellular networks to update the pre-catched contents. Using its computing facility, the Fog server can also collect and process the utility data of users, such as number of passengers on-board and their access behaviors, and reports to cloud.

\section{Comparison to Cloud Computing}

The major difference between cloud computing and Fog computing is on the support of location awareness. The cloud computing locates in a centralized place and serves as a
centralized global portal of information; cloud computing is often lack of location awareness. The Fog computing extends cloud to reside at user's premises and dedicates on localized service applications. Table~\ref{tab: compare} summarizes the
differences between Fog computing and Cloud computing.

Note that the Fog servers at different locations can be deployed by separate operators and owners, and form a collaborative Fog computing system in the wide region. For example, \cite{luan2011vtube} describes a distributed vehicular Fog computing system where Fog servers in a city are deployed by separate entities for their own commercial usage. The Fog servers deployed by different owners work in a fully distributed manner, and are formed as an integrated content distribution network for disseminating media contents to vehicles across the city. In the example of \cite{luan2011vtube}, the content files to distribute is uploaded to the Fog servers by their owners using local wireless connections.

\section{Design from Storage, Computing and Communication}

Unlike the gateway device in traditional access networks, \emph{e.g.}, WiFi and cellular network, A Fog server is a generic virtualized equipment with the on-board storage,
computing and communication capability. Therefore, a Fog server is a much more powerful and flexible device managing three-dimensional resources and can deliver more intelligent and adaptive services to users.

\subsection{Storage}

In a specific service area, a Fog server predicts the mobile user's demand
on information and pre-cache the most desirable contents accordingly using a
proactive way \cite{bacstuug2014living}. Such information can be either retrieved from
the cloud or uploaded by its owner. For example, the Fog
servers deployed in the airport can pre-cache the flight and local
transportation information which is desirable to travellers in the airport.
Therefore, the key design issue of Fog computing is to predict the user's
demand and proactively select the contents to cache in the geo-distributed
Fog servers based on the specific locations.

\subsubsection*{Compared to Related Systems}

The Content Delivery Network (CDN) \cite{peng2004cdn} represents the most
mature catch networks and extensively pursued in both academic and industry.
CDN is the Internet-based cache network by deploying cache servers at the
edge of Internet to reduce the download delay of contents from remote sites.
CDN mainly targets to serve traditional Internet users, which have
much broader interests and blur service demands that are more difficult to
predict than those of mobile users. With the precise service region, a Fog
server, in contrast to a CDN server, has much clearer target users and service demand. It is thus key for Fog servers to explore location feature to fully utilize its storage and
computing resources to provide the most desirable services to mobile users.

Information Centric Network (ICN) \cite{ahlgren2012survey} is a wireless cache infrastructure which provides content distribution services to mobile users with distributed cache
servers. Different from the cache servers in ICN, the Fog servers are
intelligent computing unit. Therefore, the Fog servers are not only used for
caching, but also as a computing infrastructure to interact with mobile
users and devices for real-time data processing. The Fog servers can be
connected to the cloud and leverage the scalable computing power
and big data tools for rich applications other than content distribution,
such as Internet of Things, vehicular communications and smart grid
applications \cite{bonomi2012fog}.

Ba{\c{s}}tu{\u{g}} \emph{et. al}. \cite{bacstuug2014living} show that the
information demand patterns of mobile users are predictable to an extent and
propose to proactively pre-cache the desirable information before users
request it. The social relations and device-to-device communications are
leveraged. Fog computing is a much more broad and generic paradigm as compared to \cite{bacstuug2014living}; the proactive caching framework described in \cite{bacstuug2014living} can be applied in Fog computing.

\subsection{Computing}

A salient feature that differentiates Fog computing from the traditional
cache networks and access technologies is that Fog servers are an intelligent computing system. This allow a Fog server to autonomously and independently serve local computation
and data processing requests from mobile users. \cite{newcloudlets} shows
the applications of Fog computing in the cognitive applications. In another
example, a Fog server inside the shopping mall or scenery park can maintain an
on-board geographic information system, and provide the real-time navigation
and video streaming to connected mobile users.

Bridging the mobile and cloud, a Fog server can also be conveniently used to
collect the environmental or utility data from mobile users at the
deployed spot, and transmit the collected big data to cloud for in-depth
data analysis; the results can be provided to third party for strategic and
valuable insights on business and government event planning, execution and
measurement.

\subsubsection*{Compared to Related Systems}

Despite of the high computing power, the cloud is faraway from mobile users
and can hardly support real-time computing intensive applications due to the
constrained bandwidth of networks. The demand of real-time
resource-intensive mobile applications, \emph{e.g.}, cognitive and
IoT applications, motivates the design of ubiquitous edge
computing system \cite{newcloudlets, zhang2006transparent}. Cloudlets \cite%
{cloudlets, newcloudlets} adopt the same framework of Fog computing, in
which a Cloudlet server, similar to the Fog server, is deployed in the
proximity of mobile users and processes the local computing requests of mobile
devices at real-time for video streaming and data processing. A comparison of processing delays using Cloudlets and Amazon clouds is shown in \texttt{http://elijah.cs.cmu.edu/demo.html}. The Cloudlets as described in \cite{cloudlets, newcloudlets} primarily focuses on providing computing services \cite{meurisch2015upgrading}; the Cloudlets, however, can be easily adapted to Fog computing. Transparent computing \cite{zhang2006transparent} is a highly virtualized system, which targets to develop the computing system transparent to users with cross-platform and cross-application support.

The Fog computing is a generic platform for edge computing and, more importantly, Fog computing focuses on serving the localized information applications and computation requests. The prototype and techniques in \cite{newcloudlets, zhang2006transparent} can be
incorporated in Fog computing framework.

\subsection{Communication}

A Fog server is an intermediate networking component which connects with
mobile users, peer Fog servers and cloud. The unique role of Fog servers raise rich
research issues from networking perspective. In addition, as Fog servers are highly
virtualized systems, it can monitor the network behaviors and adapt applications accordingly. The next section will elaborate on the research issues of Fog computing from the newtorking perspective in details.

\section{Research Issues from Networking Perspective}

\begin{figure}[t]
\centering
\includegraphics[width=.6\textwidth]{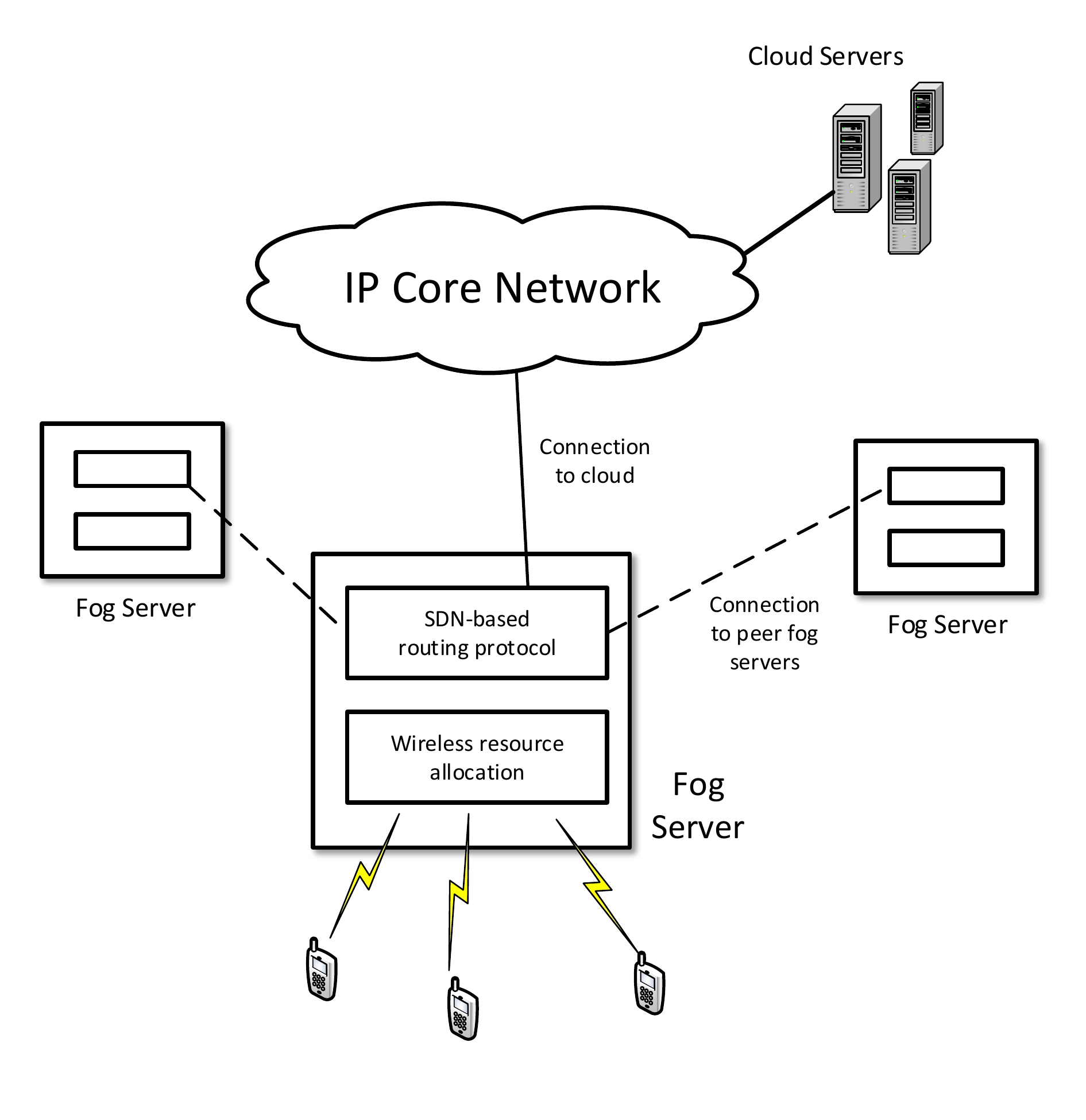}
\caption{Networking connections of Fog computing}
\label{fig: networking}
\end{figure}

With the Mobile-Fog-Cloud architecture shown in Fig.~\ref{fig: fog architecture}, a fog
server maintains three sorts of connections at the same time: wireless
connections from the Fog server to the local mobile users, wired/wireless connections among peered Fog servers, and wired/wireless connections to the cloud, as indicated in Fig.~%
\ref{fig: networking}. In this section, we discuss possible avenues of
future works from each of the three aspects.

\subsection{Communications between Mobile and Fog}

A Fog server may adopt the off-the-shelf wireless interfaces, \emph{e.g.}, WiFi and
Bluetooth, to connect with the mobile users. However, with the application- and location-awareness, the Fog server provides rich potentials
for optimal wireless resource allocation from the following two aspects:

\textbf{Cross-layer Design}: Unlike traditional WiFi access points, the fog
server manages an autonomous, all-inclusive network by providing both
service applications and wireless communications to mobile users in the
coverage. Therefore, a Fog server can manage all the communication layers
and effectively enable the cross-layer design to
provide the best service quality to users. For example, as in
\textquotedblleft BLUE" \cite{greyhound}, a Fog server can cache a number of
videos and deliver Youtube-like video streaming services to mobile users in
the proximity. In this case, based on the context, wireless channel and
video popularity information, the video services can be conveniently adapted
towards the optimal performance via cross-layer adjustifications.

\textbf{Predictable User Feature and Demand}: With specific location, mobile
users typically present predictable features and service demand. For
example, a Fog computing system deployed in the shopping mall needs to
address the diverse mobilities of users, whereas the similar system deployed
in the inter-state bus \cite{greyhound} only needs to consider static
on-board passengers. In addition, the fog computing system in a shopping
center may target to serve elastic traffic for ads and sales information
delivery, whereas that deployed on a bus may needs target to multimedia
applications with inelastic traffic. Therefore, a Fog server needs to adapt
to wireless interface to fully explore the localized user features and
service demand.

\subsection{Communications between Fog and Cloud}

The cloud performs two roles in a Fog computing system. First, the cloud is
the central controller of Fog servers deployed at different locations. With
each Fog server focusing on the service delivery to mobile users at specific
locations, the cloud manages and coordinates the geo-distributed Fog servers at different regions. Second, the cloud is the central information
depot. The Fog servers at different locations select the information
contents from the cloud and then deliver the replicas of contents from its cache
to the mobile users.

\begin{figure}[t]
\centering
\includegraphics[width=.8\textwidth]{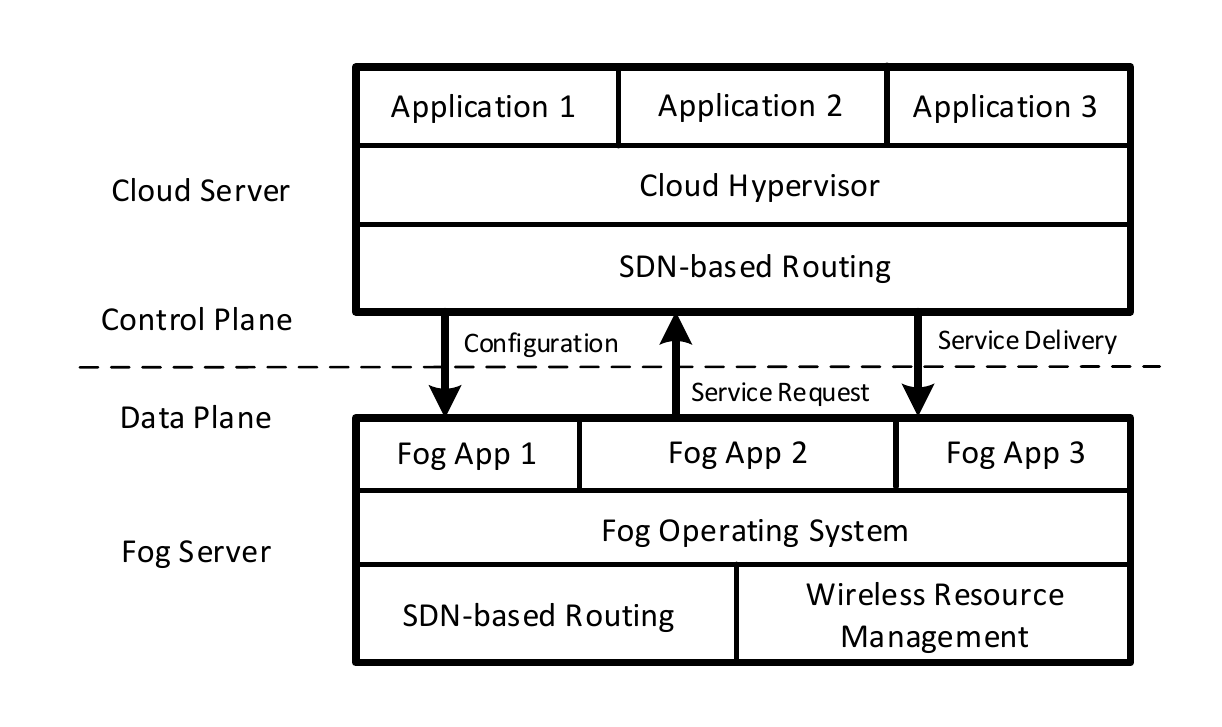}
\caption{Runtime of Fog computing system}
\label{fig: runtime}
\end{figure}

Fig.~\ref{fig: runtime} shows the software structure of a Fog computing system. A cloud server manages the applications and contents for the entire system. At a Fog server, a selective localized applications are provisioned and synchronized with the cloud.

With the dual functions of cloud, the data delivery and update from cloud to fog can be realized using a software-defined networking (SDN) approach \cite{kreutz2015software}. In this case, the traffic routing is decoupled to the control plane and data plane in which the cloud manages the network with a global view and establish the routing path of data to update the geo-distributed Fog servers.

\subsection{Communications between Fogs}

A Fog server at different locations manages a pool of resources locally. This makes the collaborative service provision and content delivery among peered Fog servers promising to improve the entire system performance. The data routing among Fog servers can be managed either by a centralized manner using the SDN-based approach, or by a fully distributed manner through the traditional routing mechanism, \emph{e.g.}, OSPF. In addition, the data transmission is challenged by the following issues:
\begin{itemize}
\item \textbf{Service policy}: as shown in Table~\ref{tab: compare} and illustrated in Fig.~\ref{fig: fog architecture}, the Fog servers at different locations may be deployed by different entities for distinct commercial usages. As a result, they may conform to different policies defined by owners and therefore the data routing among Fog servers needs to address the heterogeneous service policies.
\item \textbf{Topology}: Fog servers co-located in the same region may be connected to the Internet through the same Internet service provider with the high-rate low-cost connections, to enable efficient collaborations among nearby Fog servers can alleviate the traffic between cloud and Fog servers, and improve the system performance with saved bandwidth cost and enhanced data rate.
\item \textbf{Connection}: the data routing among Fog servers needs to consider the features of connections among Fog servers. Specifically, Fog server can be connected with each other using the wired connections over Internet or wireless connections through opportunistic connections. For example, \cite{luan2011vtube} proposes a vehicular fog computing system where contents are shipped among Fog servers by opportunistic vehicular contacts.
\end{itemize}

\subsection{Challenges of Fog Computing Deployment}

Fog computing puts additional computing and storage resources at the edge, with
the purpose to fast process the localized service requests using local
resources and connections. At different locations deployed, the Fog server, however, needs
to adapt its services, which poses extra management and maintenance cost. In addition, the network operator of a Fog computing system needs to address the following issues challenges:

\begin{itemize}
\item \textbf{Application}: At a specific location, the network operator needs to customize the applications embedded in each of the Fog servers based on the local demand.

\item \textbf{Scaling}: The network operator needs to anticipate the demand of each of the Fog servers and deploy adequate fog resources so as to sufficiently provision.

\item \textbf{Placement}: A group of Fog servers can collaboratively provide service applications to mobile users nearby. For example, multiple Fog servers can be deployed inside a shopping center to provide seamless fog applications. As such, with different user demands at different locations, how to optimally place Fog servers is challenging.
\end{itemize}

\subsection{Incorporating with Emerging Technologies}

\begin{figure}[t]
\centering
\includegraphics[width=.8\textwidth]{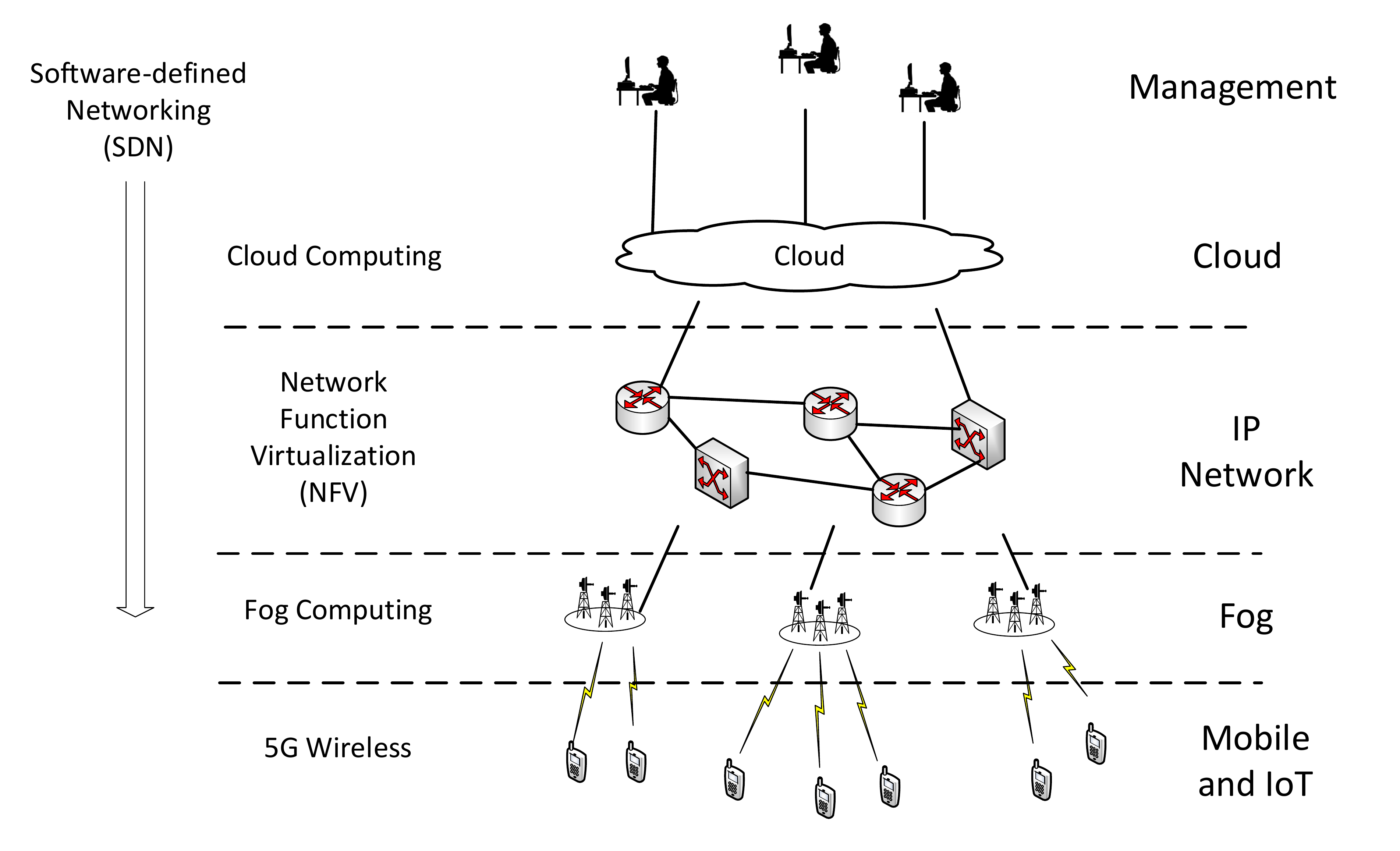}
\caption{Fog computing in emerging technologies}
\label{fig: emerging}
\end{figure}

The Fog computing can be incorporated with the emerging networking technologies with a layered architecture as shown in Fig.~\ref{fig: emerging} and described below:
\begin{itemize}
\item \textbf{5G Technologies}: Fog computing focuses on serving customized location-based applications to mobile users. The Fog layers can be adapted by using the existing accessing networks, \emph{\emph{e.g.}}, WiFi, or emerging 5G wireless technologies with a virtualized architecture as depicted in Fig.~\ref{fig: runtime}.
\item \textbf{Network Function Virtualization (NFV)}: In contrast NFV which targets to enabled virtualized network functions inside network nodes, \emph{e.g.}, switches and routers, Fog computing aims at enabling virtualized location-based applications at the edge device and providing desirable services to localized mobile users.
\item \textbf{Software-defined Networking (SDN)}: The Fog computing, as the local surrogate of cloud, needs to synchronize frequently with cloud for data update and support. With a global network view, the cloud can manage the entire network using a SDN approach.
\end{itemize}

\section{Conclusion}

This article describe Fog computing from the networking perspective. We argue that Fog computing is dedicated to serving mobile users for engaged location-based applications. By deploying reserved compute and communication resources at the proximity of users, Fog computing absorbs the intensive mobile traffic using local fast-rate connections and relieves the long back and forth data transmissions among cloud and mobile users. This significantly improves the service quality perceived by users and, more importantly,
save both the bandwidth cost and energy consumptions inside the Internet
backbone. Therefore, Fog computing represents a scalable, sustainable and
efficient solution to enable the convergence of cloud-based Internet and the
mobile computing. The purpose of this article is to investigate on the major
motivation and design goals of Fog computing from the networking
perspective. We emphasis that the emergence of Fog computing is motivated by
the predictable service demands of mobile users, and Fog computing is thus
mainly used to fulfill the service requests on localized information. As a
Fog server possesses hardware resources in three-dimensions (storage,
compute and communications), three-dimensional service-oriented resource
allocations are therefore the key of Fog computing. Moreover, with the
three-tier Mobile-Fog-Cloud architecture and rich potential applications in
both mobile networking and IoT, Fog computing also opens
broad research issues on network management, traffic engineering, big data
and novel service delivery. Therefore, we envision a bright future of Fog
computing.

\bibliographystyle{IEEEtran}
\bibliography{fog_magazine}

\end{document}